# Unveiling the Stealthy Threat: Analyzing Slow Drift GPS Spoofing Attacks for Autonomous Vehicles in Urban Environments and Enabling the Resilience

**Sagar Dasgupta***
Ph.D. Student
Department of Civil, Construction, and Environmental Engineering
The University of Alabama, Tuscaloosa, AL 35487
Email: sdasgupta@crimson.ua.edu

**Abdullah Ahmed**
Undergraduate Researcher, NSF Research Experiences for Undergraduates (REU) Site at the University of Alabama
Department of Computer Science
Ohio State University, Columbus, Ohio
Email: ahmed.898@osu.edu

**Mizanur Rahman, Ph.D.**
Assistant Professor
Department of Civil, Construction, and Environmental Engineering
The University of Alabama, Tuscaloosa, AL 35487
Email: mizan.rahman@ua.edu

**Thejesh N. Bandi, Ph.D.**
Associate Professor
Department of Physics and Astronomy
The University of Alabama, Tuscaloosa, AL 35487
Email: tbandi@ua.edu

## ABSTRACT

Autonomous vehicles (AVs) rely on the Global Positioning System (GPS) or Global Navigation Satellite Systems (GNSS) for precise (Positioning, Navigation, and Timing) PNT solutions. However, the vulnerability of GPS signals to intentional and unintended threats due to their lack of encryption and weak signal strength poses serious risks, thereby reducing the reliability of AVs. GPS spoofing is a complex and damaging attack that deceives AVs by altering GPS receivers to calculate false position and tracking information – leading to misdirection. This study explores a stealthy slow drift GPS spoofing attack, replicating the victim AV's satellite reception pattern while changing pseudo ranges to deceive the AV, particularly during turns. The attack is designed to gradually deviate from the correct route, making real-time detection challenging and jeopardizing user safety. We present a system and study methodology for constructing covert spoofing attacks on AVs, investigating the correlation between original and spoofed pseudo ranges to create effective defenses. By closely following the victim vehicle and using the same satellite signals, the attacker executes the attack precisely. Changing the pseudo ranges confuses the AV, leading it to incorrect destinations while remaining oblivious to the manipulation. The gradual deviation from the actual route further conceals the attack, hindering its swift identification. The experiments showcase a robust correlation between the original and spoofed pseudo ranges, with $R^2$ values varying between 0.99 and 1. This strong correlation facilitates effective evaluation and mitigation of spoofing signals.

**INTRODUCTION**

Autonomous vehicles (AVs) use a sophisticated combination of cutting-edge technology, including sensors, actuators, and intelligent software, to navigate roadways autonomously, thereby eliminating the need for human drivers. Their primary objective is to improve road safety, efficiency, and accessibility for all individuals. AVs show significant potential for reducing accidents caused by human error, optimizing traffic flow, and expanding transportation services to persons with driving restrictions. Society of Automotive Engineers (SAE) has developed the SAE J3016 (*1*), a classification framework spanning six levels of driving automation, to standardize the various levels of AV automation. This spectrum extends from Level 0, where human drivers retain full control, to Level 5, where autonomous vehicles can operate autonomously under certain conditions. The SAE J3016 standard provides a shared foundation for AV research, testing, and deployment, and is supported by automotive and technology-related companies worldwide. For the sake of coherence, the term "AVs" embraces vehicles with SAE Level 5 driving autonomy throughout this paper. SAE Level 5 is the pinnacle of automation under the standard, indicating complete autonomy devoid of human-operated controls such as steering wheels and pedals. This unprecedented level of automation is anticipated to change the transportation industry, bringing in a plethora of benefits like greater safety, more efficiency, and increased accessibility.

AVs rely significantly on GNSS (Global Navigation Satellite System) for exact location determination and efficient navigation from origin to destination. GNSS provides the vehicle with real-time location data, allowing it to determine its exact position on the map at any given time. This vital information is seamlessly linked with high-definition maps and other sensors, including cameras and Light Detection and Ranging (LiDar), enabling the AV to navigate highways expertly, recognize impediments, and make intelligent decisions. Multiple satellite constellations, including the well-known Global Positioning System (GPS), Globalnaya Navigazionnaya Sputnikovaya Sistema (GLONASS), BeiDou, Galileo, Quasi-Zenith Satellite System (QZSS), and Indian Regional Navigation Satellite System (IRNSS) or Navigation with Indian Constellation (NavIC), provide the extensive coverage of the Earth (*2*–*7*). In this article, we generally refer to GPS in our studies, but it also represents to GNSS where applicable. The availability of these satellite systems improves the accuracy and range of GPS, which are vital for ensuring the safety of passengers and other road users, including pedestrians and public transportation vehicles. In addition, GPS provides the AV with useful information regarding its proximity to both stationary and moving objects, including other vehicles, pedestrians, and traffic signs. This abundance of data enables the AV to make well-informed route decisions, navigate through complex road conditions, and proactively avoid potential dangers. Given its indispensability, GPS emerges as a fundamental and essential component of AV technology, playing a vital role in the future deployment of safe, efficient, and accessible autonomous vehicles.

GPS/GNSS is sensitive to different disturbances and attacks (*8*), putting at risk the precision and dependability of position and navigation information. These vulnerabilities include unintended interference (*9, 10*), such as jamming caused by walls and ceilings in garages and tunnels, multipath issues due to high-rise buildings in urban areas, atmospheric effects like scintillation and solar activity, as well as GNSS segment errors resulting from erroneous data uploads and space vehicle (SV) faults. In addition, deliberate cyber-attacks, such as jamming and spoofing, and even physical deterioration of satellite communications. Signal interference occurs when extraneous signals, such as radio transmissions or unfavorable weather conditions, interfere with GPS signals, resulting in a reduction in precision or a full loss of signal. Cyberattacks in the form of jamming are capable of interfering with the navigation systems of AVs by disrupting GPS signals. Spoofing (*11*) is a concerning cyberattack that includes creating misleading GPS signals to compromise GPS receivers, resulting in inaccurate position information. As the navigation and decision-making of AVs significantly rely on GPS signals, this type of attack can have serious consequences for AV safety. In a spoofing attack, malicious actors modify GPS signals to trick an unmanned aerial vehicle (UAV) into believing it is in a different position, resulting in navigational errors and possible safety risks(*12*). Replay and forged signal attacks are two of the most common forms of spoofing performed by attackers.

Back in 2001, the U.S. Department of Transportation acknowledged the vulnerabilities of GPS in the Volpe report (*13*). In an effort to address GPS vulnerabilities, the U.S. Department of Homeland Security (DHS) Science and Technology Directorate (S&T) has taken significant steps. They released the GPS Receiver Allow List Development Guide (*14*) and an updated version of the Positioning, Navigation, and Timing (PNT) Integrity Library (*15*). These resources serve to safeguard GPS devices from spoofing, a deceptive technique that involves manipulating GPS signals with false information. Furthermore, the federal government has issued an executive order focused on "Strengthening National Resilience Through Responsible Use of Positioning, Navigation, and Timing Services. (*16*)" This highlights the significance of a resilient GPS service to ensure safe and secure navigation on roadways. The initial step towards developing such resilience involves identifying vulnerabilities and conducting attack modeling. By simulating sophisticated attacks, the current GPS service can be tested, and from the attack patterns, detection and mitigation strategies can be formulated. This paper presents a stealthy and sophisticated GPS spoofing attack scenario wherein the attacker gradually shifts the perceived location of an AV to match its actual state. This attack serves as a foundation for creating a robust GPS. The gradual and subtle nature of the location shift makes it difficult to detect, and it may be overlooked by the AV's defense system, being mistaken as unintentional errors from onboard sensors or caused by multipath or environmental conditions.

The subsequent sections of this paper are structured as follows. The “Related Work” section provides an overview of existing spoofing attack methods and identifies the research gap. In the “Receiver Location Calculation” section, an algorithm for calculating the receiver location based on GPS navigation and observation data is presented. The “GPS Spoofing Attack Modeling” section introduces the stealthy GPS spoofing attack modeling framework. The experimental setup for the GPS spoofing attack is outlined in the “Experimental Setup” section. The “Results and Discussion” section presents the results and ensuing discussion from the conducted experiment. Lastly, “Conclusions” section offers concluding remarks and outlines potential avenues for future research.

## RELATE WORK

Spoofing attacks can be broadly categorized into two types: Asynchronous attacks and Synchronous attacks (*17*). Asynchronous attacks involve producing a spoofed signal with a different timestamp compared to the legitimate signal. One common example of an Asynchronous attack is a meacon or replay attack, where a legitimate signal from a previous timestamp is retransmitted at a later time to enhance its authenticity (*18*). On the other hand, synchronous spoofing attacks aim to generate spoofed signals with timestamps that match those of legitimate signals. These attacks require prior knowledge of the receiver's location information, making them more challenging to execute due to the difficulty in obtaining such critical details. Advanced synchronous attacks, like the nulling attack, go a step further by entirely canceling out authentic signals, creating a clear path for the uncontested transmission of spoofed signals to the target (*12*). An example of a specific type of synchronous attack is the slow-poisoning distance-decreasing attack, which falls under the category of replay attacks. This attack involves altering the transmission time of a signal to modify the pseudorange, thereby deceiving the receiver about the actual distance. In this paper, we explore relevant literature concerning distance-decreasing attacks and the various detection mechanisms employed to counter such deliberate threats.

Distance-decreasing attacks, like replay attacks, are relatively agnostic to cryptographic advancements in GNSS anti-spoofing (*18*). They rely on reducing the target's pseudorange, or perceived distance from each GPS satellite, in order to alter the final calculated position by the target’s receiver. Motallebighomi et al. (*19*) demonstrate the resistance of distance-decreasing attacks to cryptography by showing how targets can be successfully spoofed up to 4000km away from their true destination (*18*). In the study, the target was able to be spoofed within any location such that there was a set of overlapping satellites. This was done by altering the transmission time of the signals, consequently changing the perceived pseudorange of the satellite by the receiver. By changing the transmission time, the attacker need not generate their own navigation messages, thus avoiding detection by cryptographic mechanisms in place. In another study, (*20*) et al. demonstrate that a 1-millisecond change in the transmission time introduces a shift in the target receiver's perceived location of 71.89km (*21*). The study shows that changes in transmission time of the order of 1-millisecond can change the pseudorange by around 300km in an unprotected L1 C/A GPS signal. The L1 C/A GPS Signal is compatible with all GNSS devices and is optimal for consumers (*22*). The aforementioned studies all implemented distance-decreasing attacks that were successfully able to get past cryptographic security protocols. However, there has been no study that implements such an attack on a smaller magnitude that shifts the target's location slowly from its initial position.

Spoofing threats can also be classified into three distinct types: simplistic, intermediate, and sophisticated (*23*). Simplistic attacks involve the use of a commercial GPS signal simulator, coupled with a power amplifier and antenna (*24*). These attacks typically fall under the asynchronous category and resemble signal jamming, forcing the GPS receiver to lose lock and undergo partial or complete reacquisition. Such attacks are relatively easy to detect. Intermediate attacks, on the other hand, employ portable receiver spoofers. The advent of software-defined radio (SDR) technology (*25*) has made it cost-effective and straightforward to develop these portable spoofers, giving rise to intermediate-level attacks. In this scenario, the attacker closely tracks the victim AV to gain knowledge about its GPS receiver antenna's position and velocity, allowing precise positioning of the spoofed signals in relation to the legitimate signals at the AV's antenna. Since these attacks are synchronous in nature, they are considerably more challenging to detect. Sophisticated attacks represent the highest level of sophistication, utilizing multiple phase-locked portable receiver-spoofers. These attacks are designed to deceive angle-of-arrival based defensive systems, making them highly formidable and difficult to counter.

Various methods can be employed to execute attacks, but their ultimate goal is to manipulate the pseudorange calculation, leading to erroneous location calculations by the AV's GPS receiver. This study delves into the relationship between the pseudorange in legitimate and spoofed signals. By analyzing data from both legitimate and spoofed GPS signals, we determine the necessary adjustments in pseudorange to create a slow drift in the AV's perceived route compared to the legitimate route. For stealthy execution, the spoofed signals mimic the same set of GPS satellites visible to the AV receiver, while preserving the integrity of the navigation messages containing clock correction factors, ephemeris, and integrity (CEI) data. These measures ensure that the attack remains unnoticed and undetected, enabling the AV to be misled without raising any suspicion.

## RECEIVER LOCATION CALCULATION

Emulating the spoofed path necessitates the development of a software program capable of utilizing satellite signal data to calculate the receiver's location. This software program will play a pivotal role in determining the desired pseudorange values required to emulate the spoofed location. Subsequently, the simulated spoofed signals can be generated based on these calculated pseudorange values. By effectively implementing this process, the spoofed path can be generated to deceive the GPS receiver and manipulate the navigation system. Note that only GPS satellite signals and corresponding data are used for receiver location calculation in this study.

The pseudocode of the GPS receiver location calculation is presented below. The calculation is done using three steps. In the first step, the position of the satellite is calculated. Then the distance of the receiver from the satellite is calculated and finally receiver location and the receiver clock bias are calculated. GPS satellite signal data are acquired from the Receiver Independent Exchange Format (RINEX) (*26*) observation and navigation files. The navigation file contains the clock correction coefficients, ephemeris, and integrity (CEI) data as shown in **Table 1**. The observation file contains the pseudorange and carrier phase data. Pseudorange is the calculated distance between the receiver and the satellite based on the time it takes the signal to reach the receiver from the satellite and multiplying it by the speed of light (See **Equation 1**).

$$p_r^s(t_A) = c\ (t_A - t_E) \qquad (1)$$

Where, $p_r^s(t_A)$ is the pseudorange or distance between satellite(s) and receiver $r$ at time $t_A$, which stands for the time of arrival. $c$ is the speed of light and $t_E$ is the time when the signal was emitted from the satellite. As the receiver clock is not as accurate as the satellite atomic clock which introduces errors in the distance calculation, the range is called pseudorange. The receiver used correlations of the satellite PRN codes for the pseudorange calculation. The location of the GPS satellite is calculated using the ephemeris data in the navigation file and using the equations provided in (*27*)Table 7.9 of the reference (*27*). The values of constants used for the calculations are presented in **Table 2**. The location of the satellite is first determined in Earth-centered Earth-fixed (ECEF) coordinate system and then converted to the World Geodetic System (WGS) 84. Satellite (SV) clock bias is also corrected using the SVclockBias, SVclockDrift, and SVclockDriftRate data. To accurately calculate the distance of the satellite, the receiver Ionospheric and Tropospheric corrections are also performed. Least square regression is used to solve for the receiver location from satellite positions. At least data from four satellite signals data are required to solve for four unknowns i.e., user latitude, longitude, altitude and the receiver clock bias.

**Algorithm 1** Receiver Location Calculation

```
xu ← [0, 0, 0]                                          ▷ initial estimate for user position
b ← 0                                                   ▷ initial estimate for user clock bias
SV s ← all observed satellites
for sat in SVs do
    dsv ← satellite clock bias(sat)
    C ← Speed of Light
    sat, pseudorange ← sat.pseudorange + C ∗ dsv − C ∗ sat.T gd
end for
Xs ← [ ]                                                ▷ satellite position matrix
Pr ← [ ]                                                ▷ pseudorange matrix
while norm(dx) > 0.1 and norm(db) > 1 do
    for sat in SVs do
        cpr ← sat.pseudorange − b                       ▷ corrected pseudorange
        xs ← satellite position(sat)
        Xs append xs
        Xs append xs
    Xs append xs end for
    x , b ← estimate position(Xs, Pr, length(SV s), xu, b)
    dx ← x_ − xu
    db ← b_ − b
    xu ← x_
    b ← b_
end while
return xu, b
```

**TABLE 1 CEI Variables from RINEX Navigation File**

| Measurements | Description | Sample Data |
|---|---|---|
| Satellite ID | Satellite system (G) number (PRN) | G18 |
| Time | GPS Clock Time | 7/14/23 22:00 |
| SVclockBias | Space Vehicle clock bias correction coefficient | -3.46E-04 |
| SVclockDrift | Space Vehicle clock drift correction coefficient | -1.42E-11 |
| SVclockDriftRate | Space Vehicle clock drift rate correction coefficient index | 0.00E+00 |
| IODE | Issue number of the satellite ephemeris data set, Issue of Data, Ephemeris | 1.46E+02 |
| Crs | Amplitude of the Sine Correction Term to the Orbit Radius | -5.00E+01 |
| DeltaN | Mean Motion Difference from Computed Value at Reference Time | 4.37E-09 |
| M0 | Mean Anomaly at Reference Time | -3.71E-01 |
| Cuc | Amplitude of Cosine Harmonic Correction Term to the Argument of Latitude | -2.54E-06 |
| Eccentricity | Eccentricity | 3.45E-03 |
| Cus | Amplitude of Sine Harmonic Correction Term to the Argument of Latitude | 1.20E-06 |
| sqrtA | Square root of the semimajor axis | 5.15E+03 |
| Toe | Time of Ephemeris | 5.11E+05 |
| Cic | Amplitude of the Cosine Harmonic Correction Term to the Angle of Inclination | 1.30E-08 |
| Omega0 | Longitude of Ascending Node of Orbit Plane at Weekly Epoch | 1.52E+00 |
| Cis | Amplitude of the Sine Harmonic Correction Term to the Angle of Inclination | -7.45E-09 |
| i0 | Inclination Angle at Reference Time | 9.74E-01 |
| Crc | Amplitude of the Cosine Harmonic Correction Term to the Orbit Radius | 3.60E+02 |
| omega | Argument of Perigee | 3.13E+00 |
| OmegaDOT | Rate of Right Ascension | -8.27E-09 |
| IDOT | Rate of Inclination Angle | 5.04E-11 |
| CodesL2 | Codes on L2 channel | 1.00E+00 |
| GPSWeek | GPS week number | 2.27E+03 |
| L2PFlag | L2P data flag | 0.00E+00 |
| Svacc | Satellite Vehicle Accuracy | 2.00E+00 |
| health | Satellite health | 0.00E+00 |
| TGD | Group Delay Differential | -8.38E-09 |
| IODC | Issue number of the satellite clock data set | 4.02E+02 |
| TransmissionTime | Transmission time of the message | 5.04E+05 |

**TABLE 2 List of Constants and Values Used for Receiver Location Calculation**

| Constant | Details | Value | Unit |
|---|---|---|---|
| $c$ | Speed of light | 299792458 | $\mathrm{m/s}$ |
| $\mu$ | WGS 84 value of the Earth's gravitational constant | $3.986005 \times 10^{14}$ | $\mathrm{m^3/s^2}$ |
| $\dot{\Omega}_e$ | WGS 84 value of the Earth's rotation rate | $7.2921151467 \times 10^{-5}$ | $\mathrm{rad/s}$ |
| $\pi$ | | 3.1415926535898 | |
| $R$ | Radius of Earth | 6372.8 | km |

The performance of the receiver location calculation software is checked against the ground truth location. As shown in **Figure 1**, our research team has collected GPS signal data using a NovAtel CPT7700 integrated receiver with TerraStar corrections at the University of Alabama (UA) campus roadway network. This receiver has a localization accuracy of 2.5 cm. Data was stored in the device and then converted to RINEX format. The RINEX files are then used for the receiver location calculation. The distribution of error which is the distance between the location calculated by the presented algorithm and the location solution from the NovAtel receiver is presented in **Figure 2**. The distance or error is calculated using the Haversine formula (See **Equation 2**).

$$d = 2r \sin^{-1}\left(\sqrt{\sin^2\left(\frac{\varphi_2 - \varphi_1}{2}\right) + \cos(\varphi_1)\cos(\varphi_2)\sin^2\left(\frac{\psi_2 - \psi_1}{2}\right)}\right) \qquad (2)$$

where the distance, d, is described as the shortest distance between 2 points on a sphere represented as geodetic (lat/lon) coordinates, $(\varphi_1, \psi_1)$ and $(\varphi_2, \psi_1)$. The radius, *r*, is the radius of the earth (6372.8 km). The distribution is right-skewed or positively skewed i.e., most of the error values are concentrated in the lower values. The most populated bins have errors of less than 6 m. There are fewer occurrences of error more than 18 m. The minimum and the maximum error are 0.18 m and 36.85 m, respectively. The mean and median values are 7.40 m and 6.17 m respectively. Note that location data are calculated without considering real-time kinematics correction, which is included in built-in NovAtel CPT7700, and lead to higher location error. However, **Figure 3** proves how closely the receiver location calculation algorithm matches with the ground truth path which proves its ability to be used for detecting and generating spoofed routes. The x-axis shows the latitude (lat) and the y-axis shows the longitude (lon). These results show that the presented algorithm has the capability for estimating the location of the receiver and using it for further emulating the spoofing attacks.

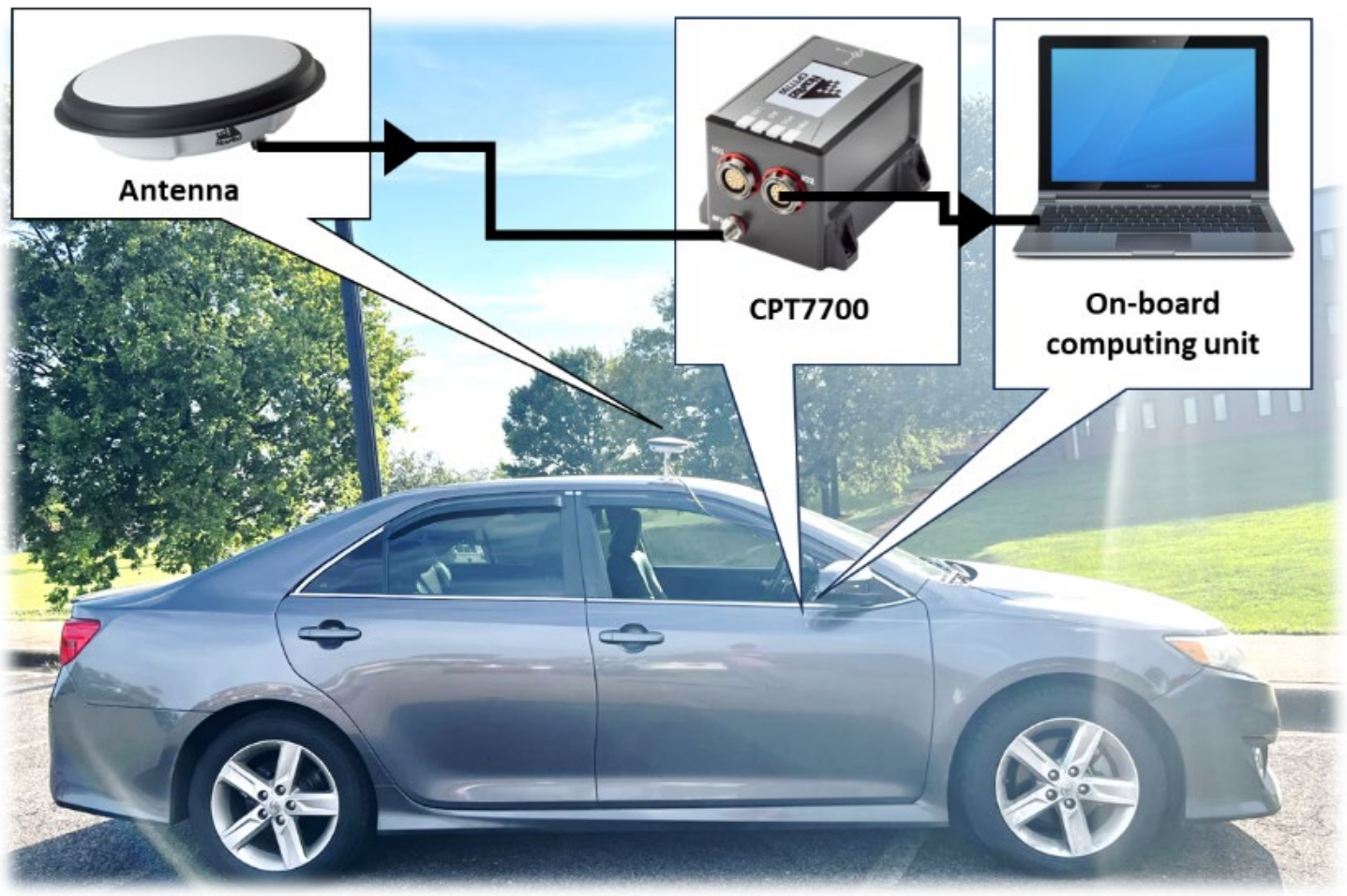

**Figure 1 Vehicle with a NovAtel CPT7700 integrated receiver with TerraStar corrections**

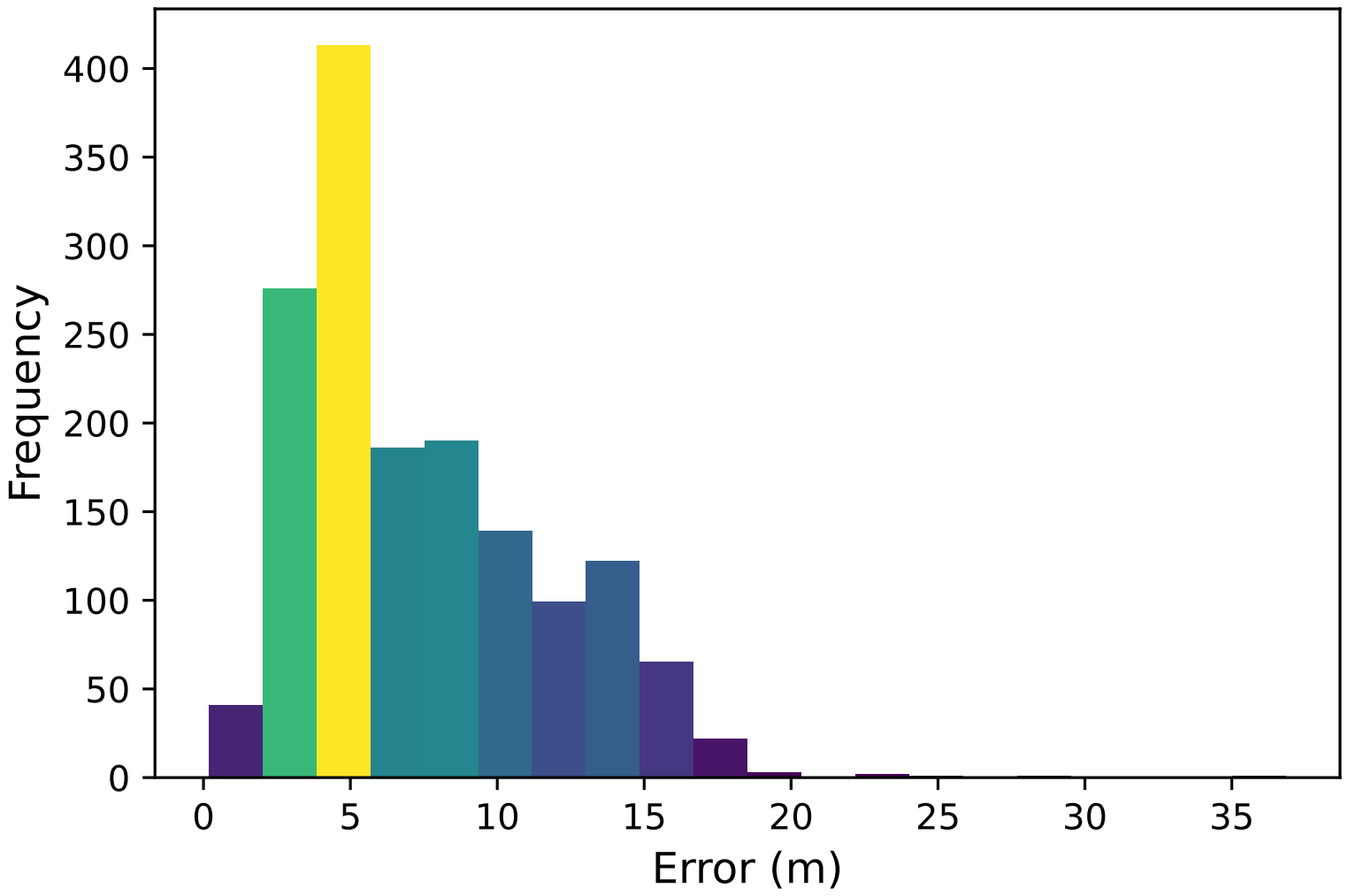

**Figure 2 Location calculation error distribution**

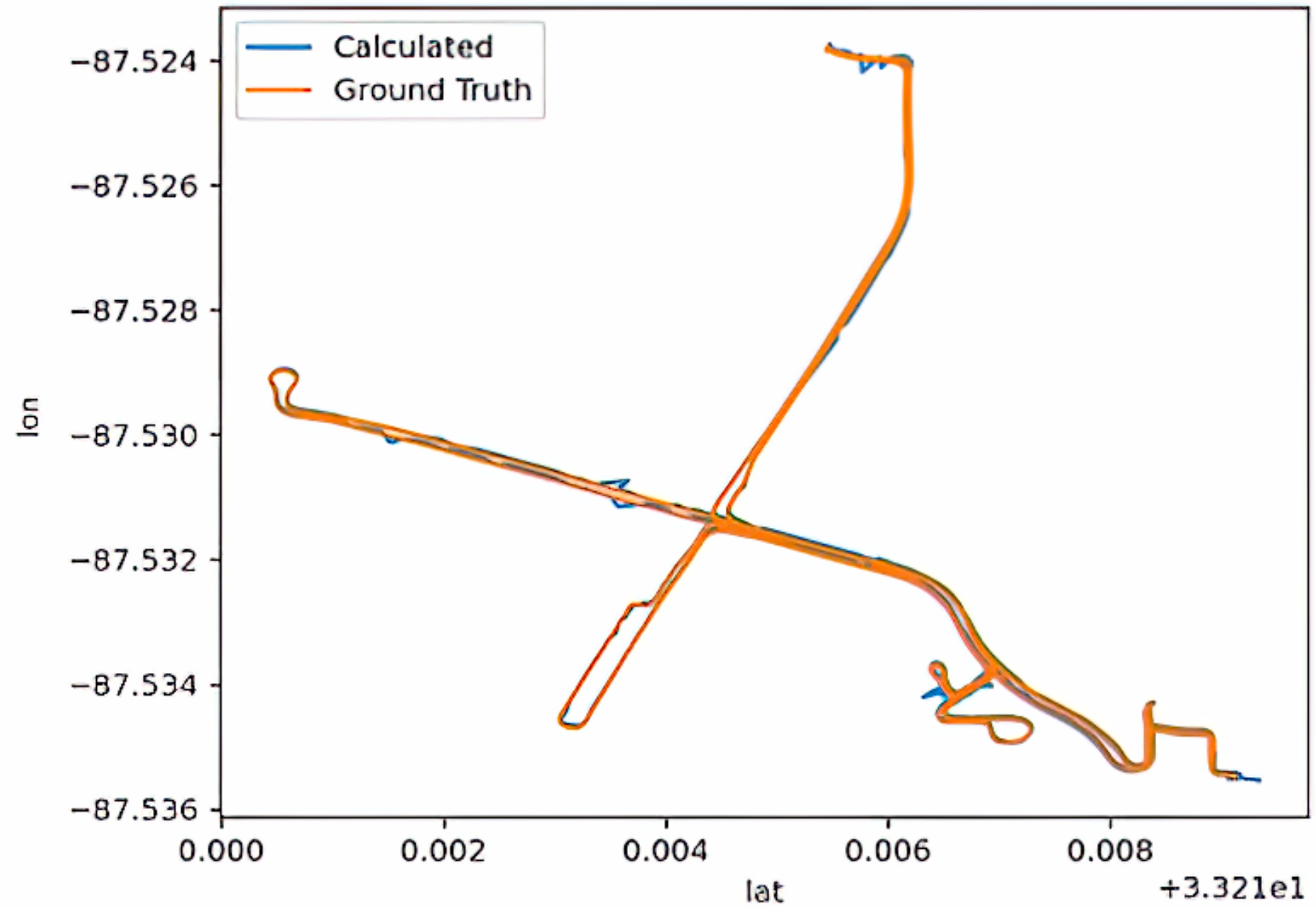

**Figure 3 Ground truth and calculated path**

## GPS SPOOFING ATTACK MODELING

The schematic representation of the slow drift stealthy attack framework is illustrated in **Figure 4**. The attacker initiates the attack armed with prior knowledge of the spoofed path locations. Throughout the attack, the attacker closely tracks the AV to ensure that both the AV and the attacker can observe the same satellites. The attacker's receiver captures legitimate GPS signals and records the navigation and observation data on the local computer. Navigation data comprises information sourced from the RINEX navigation file, while observation data includes the pseudorange information calculated by the attacker's receiver. To generate the desired spoofed pseudorange values, the true pseudorange data extracted from the observation data is combined with the pre-defined spoofed location and fed into the Spoofed Pseudorange Emulator (SPE). The SPE employs experiment-derived correlations or machine learning models to estimate pseudoranges that would yield the spoofed location. Subsequently, the calculated spoofed pseudoranges, along with legitimate CEI (Clock, Ephemeris, and Integrity) variables from the navigation data, are fed into the GPS receiver location calculation algorithm to determine the spoofed location. The calculated spoofed location is then compared to the desired spoofed location. If they match, the spoofed pseudoranges are transmitted to the Spoofer GPS Signal Simulator. In this phase, the GPS signals are manipulated by altering the signal generation time and carrier phase, ensuring that the AV receiver's localization solution yields the spoofed location. Moreover, the spoofed signal mirrors the satellites visible in the legitimate signal, exhibiting a seamless transition with no abrupt changes in signal or message properties after the attacker locks onto the AV's receiver. The pre-determined frequency of the spoofed location changes is designed to mimic a gradual drift from the AV's actual route. Finally, the simulated GPS signals are transmitted to the AV's GPS receiver, effectively executing the slow drift stealthy attack and detection.

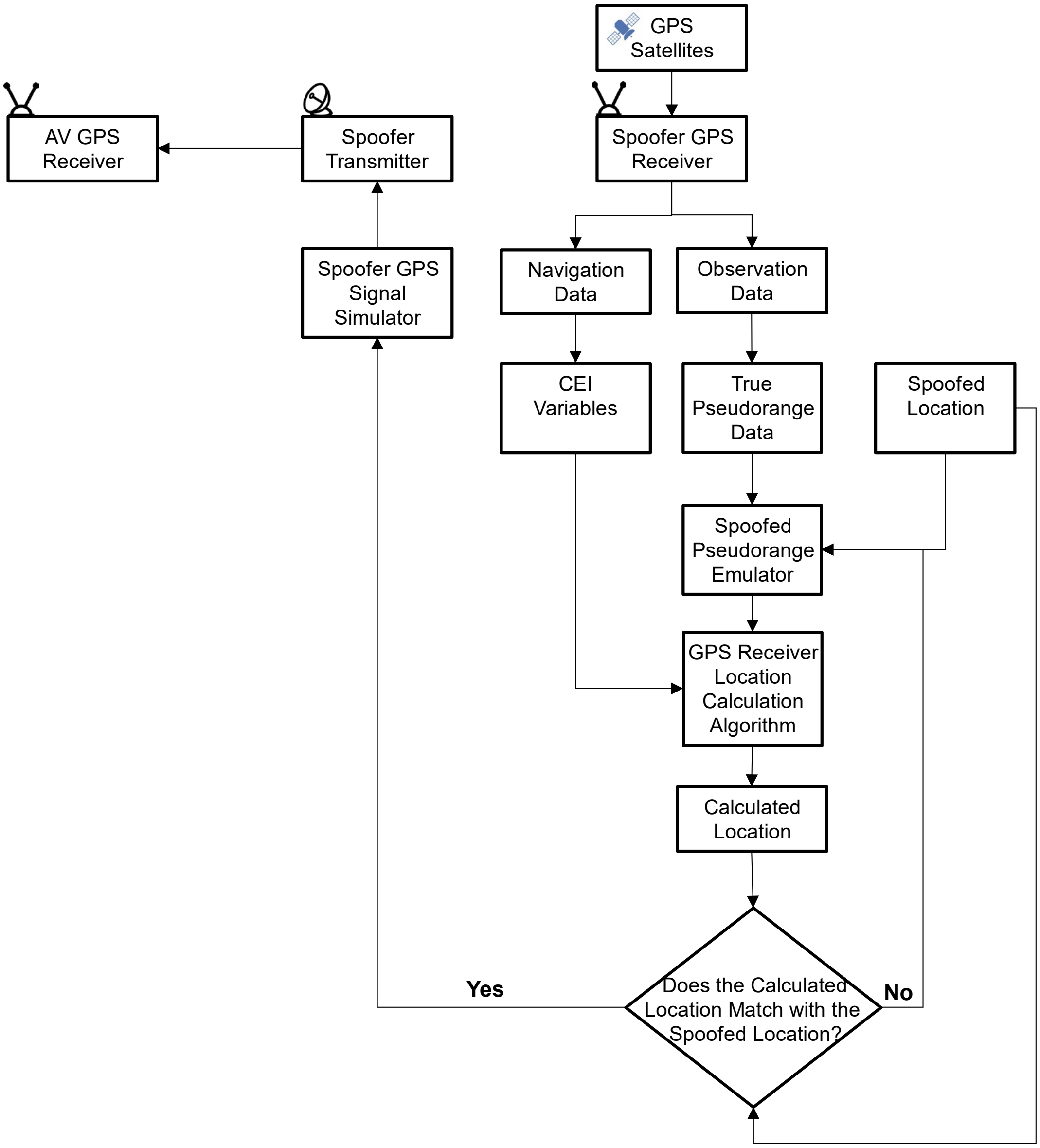

**Figure 4 Attack modeling framework**

## EXPERIMENTAL SETUP

The experimental setup for creating the dataset for creating GPS spoofing attack takes place on the premises of the University of Alabama campus. To simulate real-world conditions, a moving vehicle is used as the target, equipped with a GPS receiver as shown in **Figure 1**. Specifically, a NovAtel CPT7700 integrated receiver with TerraStar corrections is employed for this purpose. This receiver boasts an OEM7700 receiver from Hexagon-NovAtel, complemented by a high-performing Honeywell HG4930 Micro Electromechanical System (MEMS) Inertial Measurement Unit (IMU) and a VEXXIS® GNSS-800 Series Antenna. With the capability to track multi-GNSS signals across all frequencies simultaneously, this receiver is well-suited for the experiments. The localization solution output frequency is set at 1 Hz, and the receiver exhibits a high localization accuracy of 2.5 cm. After conducting the experiments, the data are converted to the RINEX format using the NovAtel application suite. As shown in **Figure 5**, The driving route carefully replicates an urban road structure, aiming to imitate real-world scenarios. To effectively assess the AV's navigational performance, the route incorporates three basic maneuvers, including going straight, taking a left turn, and taking a right turn. Ensuring the integrity of the experimental setup, the visibility of the same set of satellites is maintained throughout the entire duration of the experiment. To capture comprehensive GPS data for the entire route, all relevant information is diligently stored in the local storage system. This robust experimental setup lays the foundation for analyzing the effects of the slow drift stealthy GPS spoofing attack on the AV's navigation performance. By closely monitoring the AV's reaction to the manipulated GPS signals, valuable insights can be gained, enabling the development of effective defense mechanisms against such sophisticated attacks in the future.

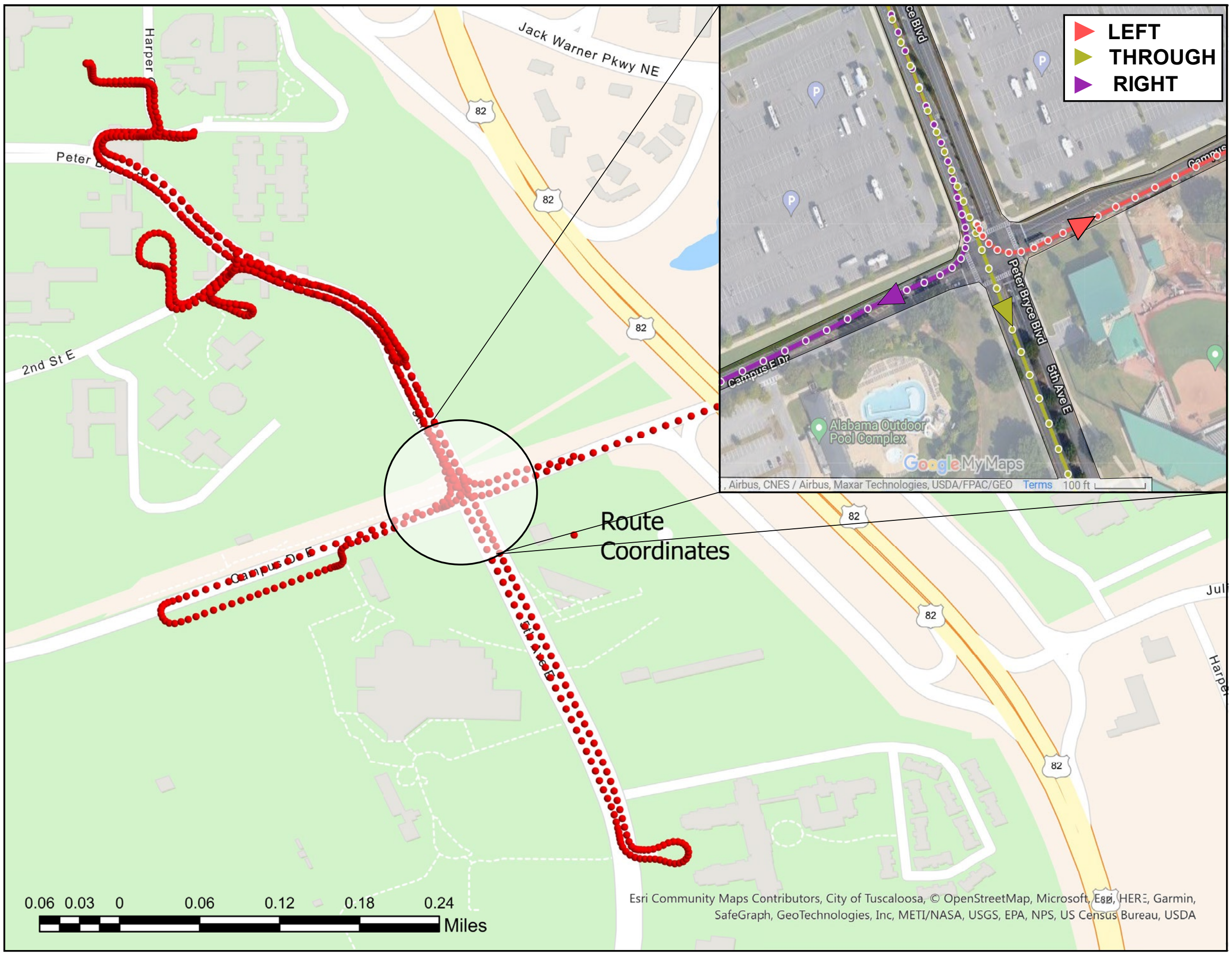

**Figure 5 Driving route during the field experiment at the University of Alabama Campus, AL**

## RESULTS AND DISCUSSION

A particular intersection from the experiment's driving route is designated for in-depth analysis. Two trajectories are singled out for examination: the right turn trajectory is treated as the legitimate route, whereas the left turn trajectory is labeled as the spoofed route (See **Figure 6**). The visible GPS satellites for both routes are G05, G10, G13, G15, G18, G23, G24, and G29. Since the number of visible satellites exceeds the minimum requirement of four GPS satellites, it is suitable for location determination. **Figure 7** presents the legitimate and spoofed pseudorange values for each visible satellite and **Figure 8** shows the difference between the legitimate and spoofed pseudorange. For G10, G18, G23, G24 the legitimate pseudorange is higher than the spoofed, and G05, G13, G15, G29 legitimate pseudorange is lower than the spoofed. Hence, for spoofed pseudorange of all the visible satellites will not be always higher or lower than the legitimate one which makes it hard to predict the spoofed pseudorange and needs multiple iterations to determine the correct set of spoofed pseudoranges for mimicking the spoofed location. We are doing further studies to understand these differences. Based on the experimental data, the correlation between the legitimate and the spoofed pseudorange for each satellite is established and shown in **Figure 9**. The $R^2$ values vary between 0.99 and 1 showing the response variable (legitimate) can be accurately explained by the predicted variable (spoofed).

This outcome shows that it is possible to emulate and/or estimate the desired spoofed pseudorange values by extracting legitimate pseudorange data from the observation data and combining it with the pre-defined spoofed location, which can be fed into the Spoofed Pseudorange Emulator (SPE). Using these calculated spoofed pseudoranges and legitimate CEI (Clock, Ephemeris, and Integrity) variables from the navigation data, one can determine the spoofed GPS receiver location to slowly guide an AV towards a desired location as per spoofer intention. Especially it is true for urban structured road network systems. As the spoofed signal mirrors the satellites visible in the legitimate signal, exhibiting a seamless transition with no abrupt changes in signal or message properties after the attacker locks onto the AV's receiver. This pre-determined frequency of the spoofed location changes can be designed to mimic a gradual drift from the AV's actual route.

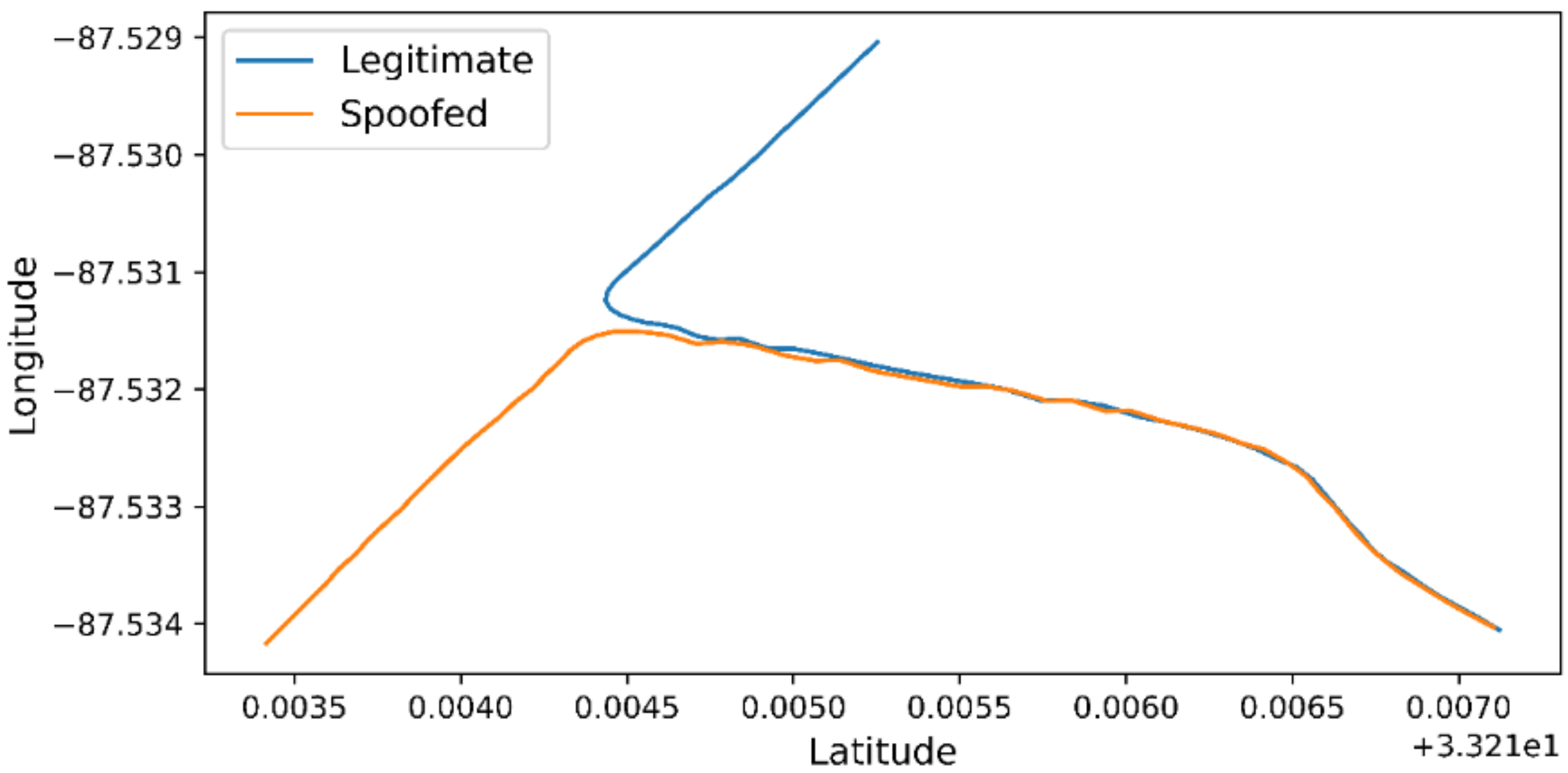

**Figure 6 Legitimate and spoofed route**

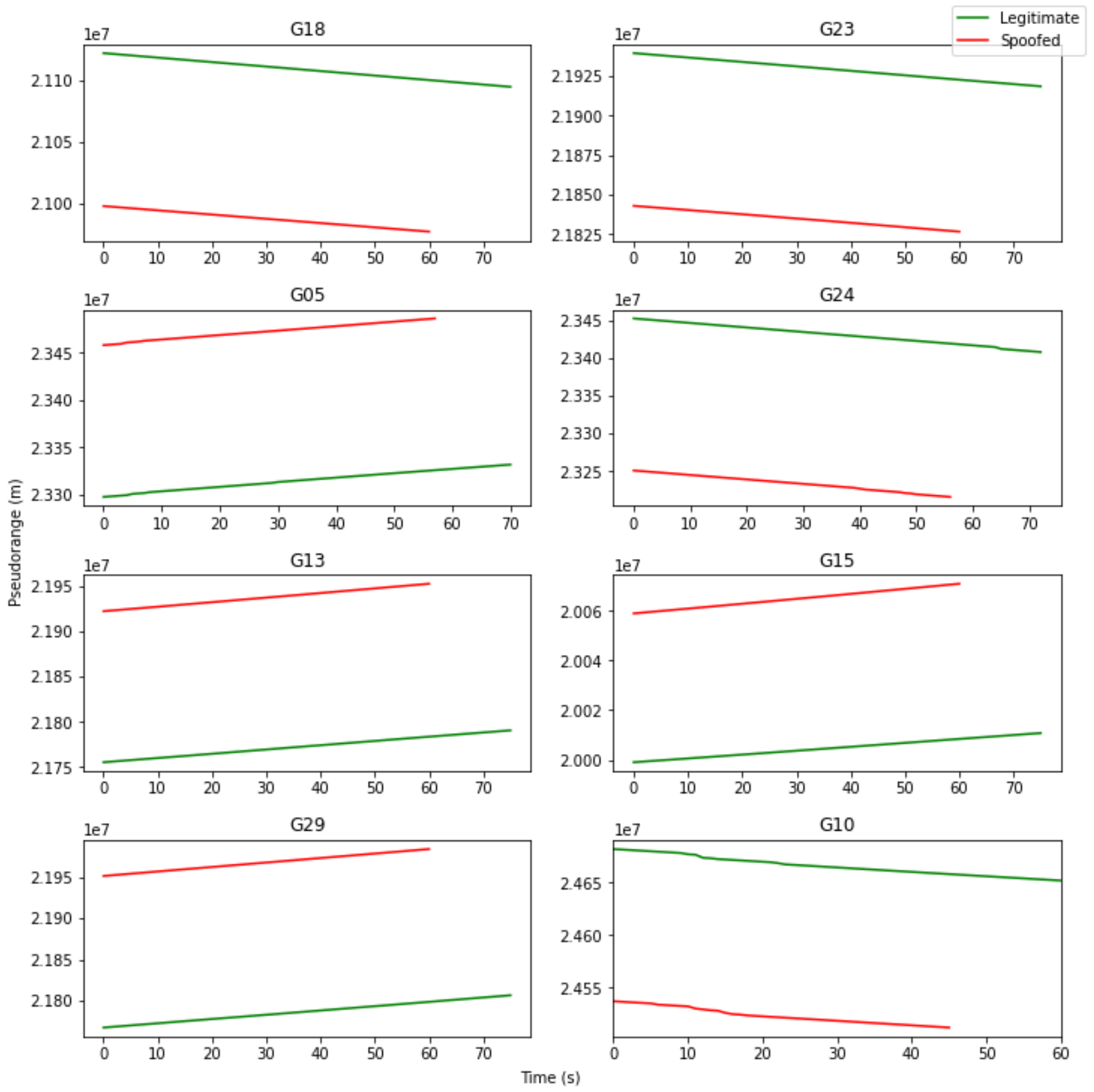

**Figure 7 Legitimate and spoofed signal pseudorange**

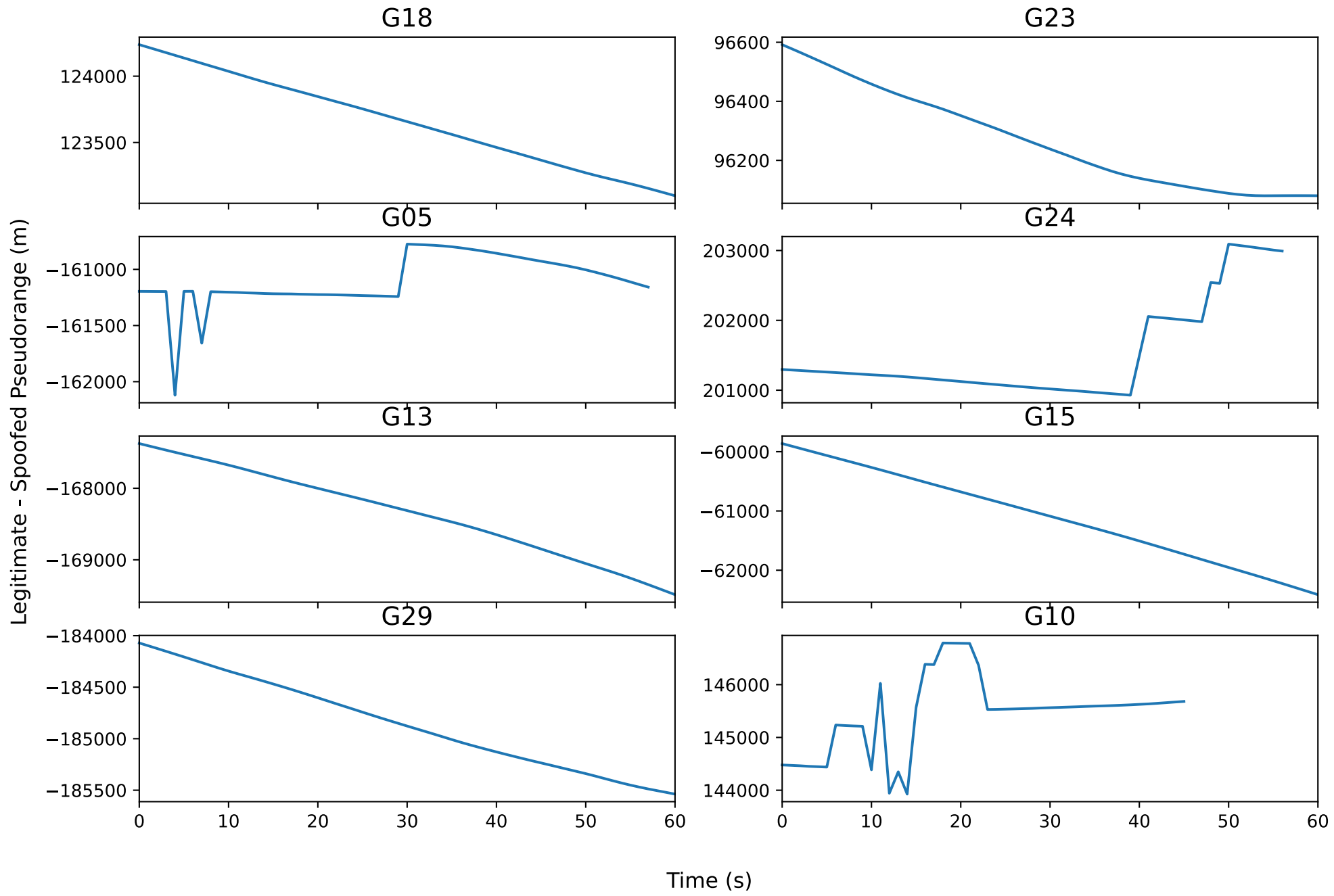

**Figure 8 Difference between legitimate and spoofed signal pseudorange**

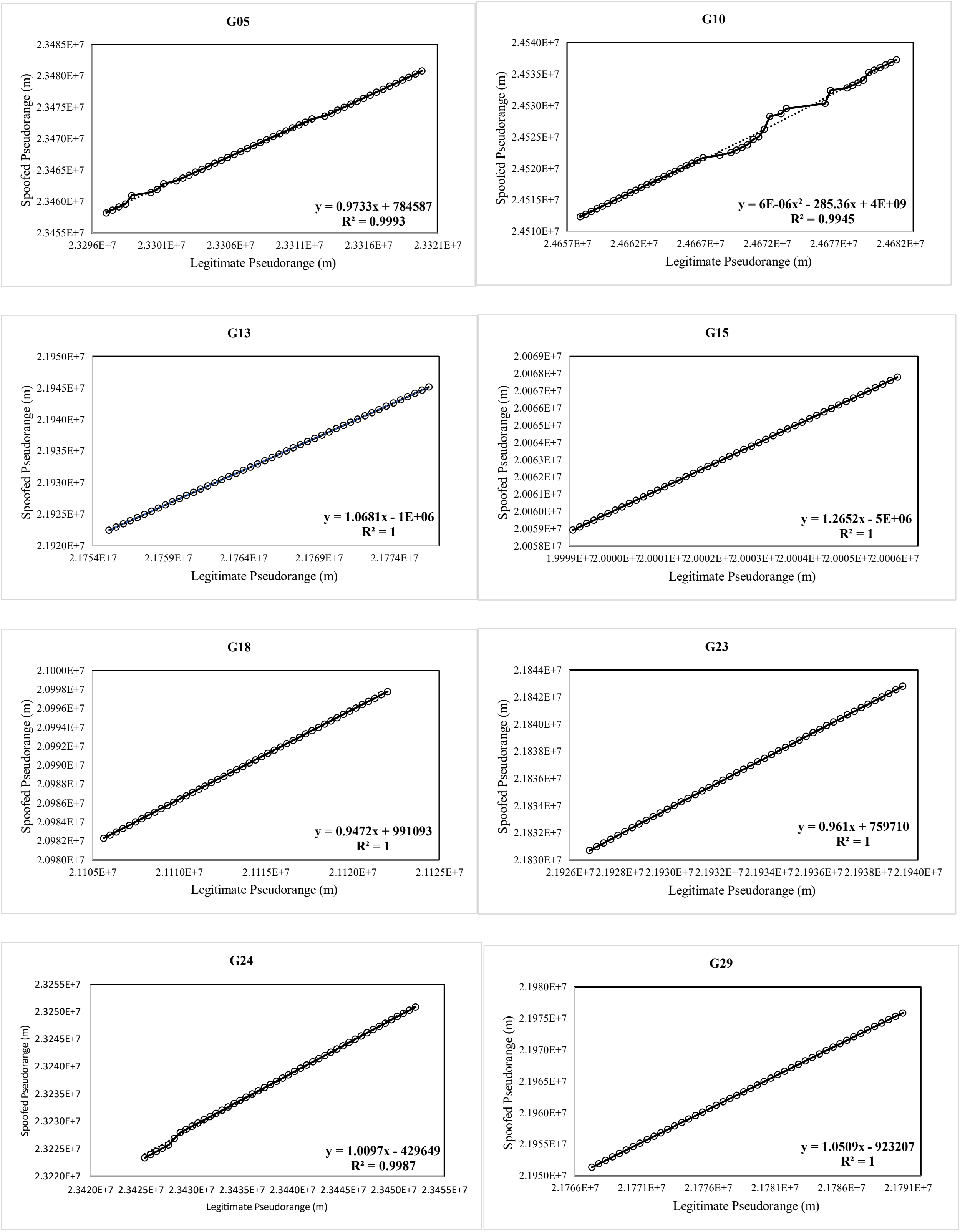

**Figure 9 Correlation between legitimate and emulated spoofed pseudorange**

## CONCLUSIONS

This study sheds light on the critical issue of GPS spoofing and its potential impact on AVs. The implication of this work can be extended to Cybersecure systems, Ariel Vehicles, and scenarios where GPS or GNSS signals are vulnerable to attacks. GPS/GNSS spoofing, in particular, emerges as a sophisticated and damaging attack, wherein malicious actors manipulate the end user receivers to provide false location information, leading to misdirection and potential safety hazards. Our research focused on a stealthy slow drift GPS spoofing attack, where we carefully emulated the satellite reception pattern of the victim AV while altering pseudo ranges to deceive the AV even during turns. The gradual deviation from the actual route added a layer of obscurity to the attack, making it challenging to identify promptly. By progressively deviating from the legitimate route, we designed the attack to evade rapid detection, posing significant risks to user safety and security. We presented a comprehensive approach for constructing covert spoofing attacks on AVs, employing the same satellite signals as the victim AV for precise execution. The manipulation of pseudo ranges effectively misled the AV without triggering suspicion. The field experiments and corresponding analyses conducted in this study demonstrated the essential correlation between original and spoofed pseudoranges, a key factor in generating convincing spoofing signals. Understanding this correlation is vital for developing effective defense mechanisms to protect AVs from such sophisticated attacks. As the transportation industry continues to embrace AV technology, it is crucial to address the security implications of GPS vulnerabilities. Our findings contribute to the ongoing efforts in fortifying AVs against GPS spoofing and enhancing their resilience to potential attacks. Moving forward, further research and innovation in anti-spoofing technologies will be essential to safeguard AVs from evolving threats. Strengthening the security of GPS services and implementing robust authentication measures will be instrumental in ensuring the safe and reliable deployment of autonomous vehicles in the future. In conclusion, this study serves as a stepping-stone towards a safer and more secure future of autonomous transportation, where AVs can navigate with confidence and resilience, free from the risks posed by GPS/GNSS spoofing attacks.

## ACKNOWLEDGMENTS

This material is based on a study supported by National Science Foundation (NSF)— Award# 2244371 and Award # 2104999. Any opinions, findings, and conclusions, or recommendations expressed in this material are those of the author(s) and do not necessarily reflect the views of the NSF, and the NSF assumes no liability for the contents or use thereof.

## AUTHOR CONTRIBUTIONS

The authors confirm their contribution to the paper as follows: study conception and design: S. Dasgupta, A. Ahmed, M. Rahman, and T. Bandi; data collection: S. Dasgupta, and A. Ahmed; interpretation of results: S. Dasgupta, A. Ahmed, and M. Rahman; draft manuscript preparation: S. Dasgupta, A. Ahmed, and M. Rahman. All authors reviewed the results and approved the final version of the manuscript.